# Beyond Binary Computers: How To Implement Multi-Switch Computer Hardware and Software and; The Advantage of a Multi-Switched Computer

Givon Zirkind
Givon Zirkind
B.Sc., Computer Science, Touro College; M.Sc. Computer Science, Fairleigh Dickinson University;
Mt. Sinai School of Medicine, Post Doctoral Researcher
givonz@hotmail.com

## ABSTRACT

This paper explores the possibilities of using a computing methodology—hardware and software—that employs technology other than binary. I refer to this as "supra-binary" computing. Software constructs that use more than binary techniques are discussed. The gains in supra-binary software are demonstrated, which includes supra-binary code being RISC. Possible hardware implementations of a computer with other than binary based architecture are demonstrated and considered. The advantages and possible disadvantages of these hardware implementations are discussed. The gain in computing speed is evaluated and demonstrated. Supra-binary processing would streamline parallel processing and make its implementation a built-in feature of the software and hardware. Also, supra-binary would bring an advancement to neural networking. This is discussed and demonstrated. In addition, possible applications of supra-binary computing to database and neural networking are discussed. Also, the possible implementations could be applied to telecommunications with dramatic results.

## Categories and Subject Descriptors

C.4. [**Computer Systems Organization**]: Performance Of Systems – *Performance attributes*
D.2.11 [**Software**]: Software Architectures – *Data abstraction*
D.2.12 [**Software**]: Interoperability – *Data mapping*
E.4 [**Data**]: Coding And Information Theory – *Data compaction and compression*
E.4 [**Data**]: Coding And Information Theory – *Nonsecret encoding schemes*
F.2.0 [**Theory of Computation**]: Analysis Of Algorithms And Problem Complexity – *General*
F.2.1 [**Theory of Computation**]: Numerical Algorithms and Problems – *Computation of transforms (e.g., fast Fourier transform)*
F.2.1 [**Theory of Computation**]: Numerical Algorithms and Problems – *Computations on matrices*
F.2.2 [**Theory of Computation**]: Nonnumerical Algorithms and Problems – *Pattern matching*
F.2.m [**Theory of Computation**]: Miscellaneous
H.0 [**Information Systems**]: General
H.1.1 [**Information Systems**]: Systems and Information Theory – *Value of information*
H.1.2 [**Information Systems**]: User/Machine Systems – *Human factors*
H.1.2 [**Information Systems**]: User/Machine Systems – *Human information processing*
H.1.2 [**Information Systems**]: User/Machine Systems – Miscellaneous
H.2.8 [**Information Systems**]: Database Management -- Database Applications -- *Image databases*
H.3.4 [**Information Systems**]: Information Storage And Retrieval – Systems and Software – *Performance evaluation (efficiency and effectiveness)*
H.3.7 [**Information Systems**]: Digital Libraries – *User issues*
I.3.4 [**Computing Methodologies**]: Computer Graphics – Graphics Utilities – *Application packages*
I.3.4 [**Computing Methodologies**]: Computer Graphics – Graphics Utilities – *Graphics packages*
I.3.6 [**Computing Methodologies**]: Methodology and Techniques – *Device independence*
I.3.6 [**Computing Methodologies**]: Methodology and Techniques – *Graphics data structures and data types*
I.3.6 [**Computing Methodologies**]: Methodology and Techniques – *Standards*
I.4.0 [**Image Processing And Computer Vision**]: General – *Image processing software*
I.4.2 [**Image Processing And Computer Vision**]: Compression (Coding) – *Exact coding*
I.4.3 [**Image Processing And Computer Vision**]: Enhancement – *Filtering*
I.4.3 [**Image Processing And Computer Vision**]: Enhancement – *Grayscale manipulation*
I.4.3 [**Image Processing And Computer Vision**]: Enhancement – *Sharpening and deblurring*
I.4.4 [**Image Processing And Computer Vision**]: Restoration – *Inverse filtering*
I.4.6 [**Image Processing And Computer Vision**]: Segmentation – *Edge and feature detection*
I.4.6 [**Image Processing And Computer Vision**]: Segmentation – *Pixel classification*
J.1 [**Computer Applications**]: Administrative Data Processing – *Law*

K.4 [**Computing Milieux**]: Computers And Society – Public Policy Issues – *Ethics*
K.4 [**Computing Milieux**]: Computers And Society – Public Policy Issues – *Transborder data flow*
K.4.m [**Computing Milieux**]: Computers And Society – Miscellaneous

**General Terms**
.

## 1. INTRODUCTION

They say that when one gives, one gets. For the longest time, I have contemplated an intrinsic shortcoming of our computers. All our computers are binary. All our computing methodology and architecture use binary exclusively. During an initial conversation, while volunteering computing services and data analysis to the Dept. of Neuroscience of Mt. Sinai School of Medicine, contemplating the computational issues and functioning of the neurons, these solutions came to me.

Switching in computers is done by toggling on/off switches. Usually, this is done electronically with 2 separate voltage levels. [MAL001] Hence, all our coding, encoding, all our programming constructs and logic—no matter how we conceptualize them, are reduced somehow to yes/no, if…then, on/off.

This is limiting.

Wouldn't it be grand, if we could select one of any number of options and choose a different path, different course of action, based upon each option?! Yes, we have switch statements and neural nets. But, their underlying implementations are always in a binary sequential mode. The equivalent of numerous sequential linear if…then statements. Our programming languages can branch to only one location.

What would the possibilities be, if we could branch to several locations (in programming code) at once? Or, if we could arrive simultaneously at the same place (in programming code) from several locations (in programming code)?

All data structures—stacks, ring buffers, pipes, etc.—are just conceptions. So too, any mimickry of branching to more than one point at a time. Ex. The "switch" statement in 'C' is merely a collection of if…then statements. The conceptualization of branching down many branches at once, for a neural net, is merely a conception—not a reality. My goal is to describe a system that can actually physically branch to multiple locations at once.

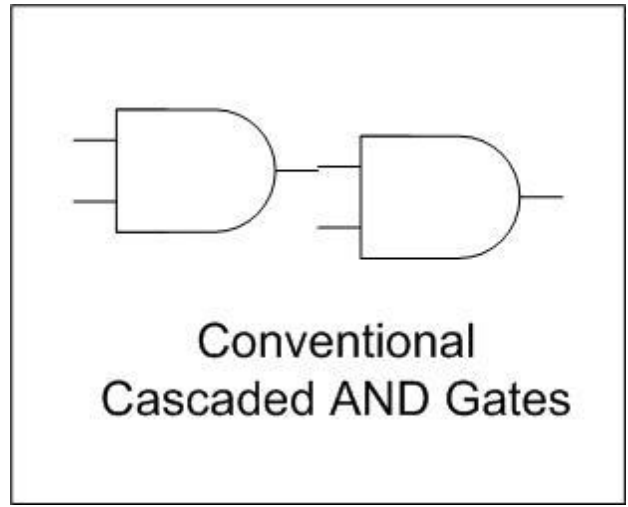

Figure 1.

In order to consider the result of a logical AND of 3 inputs, currently, we need to cascade logic circuits, as above. Likewise, programming statements are also logically ANDed by cascading.

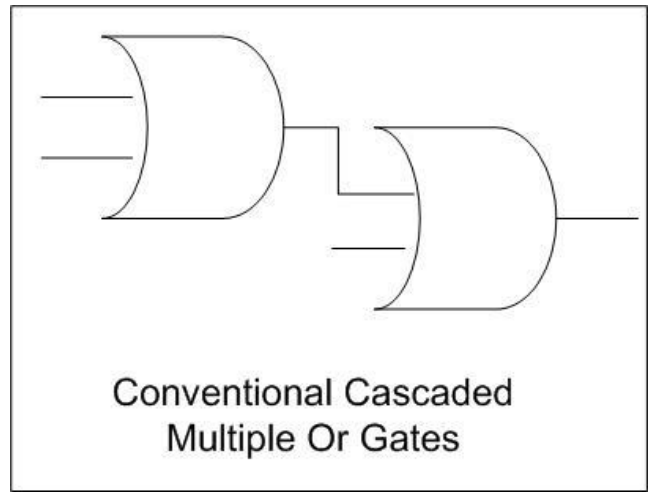

Figure 2.

The same is true of logically ORing. Both in hardware and software, a cascade is necessary.

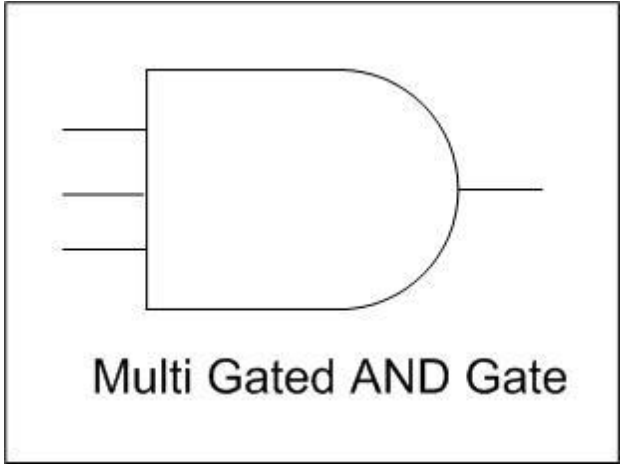

Figure 3.

What I am proposing is muli-gated AND gate as depicted above. Not a cascade of of AND gates.

| P | Q | R | P∧Q∧R |
|---|---|---|-------|
|   |   |   |       |
| T | T | T | T     |
| T | T | F | F     |
| T | F | T | F     |
| F | T | T | F     |
| T | F | F | F     |
| F | T | F | F     |
| F | F | T | F     |
| F | F | F | F     |

Table 1. The corresponding Truth Table for this multi-swtiched AND gate.

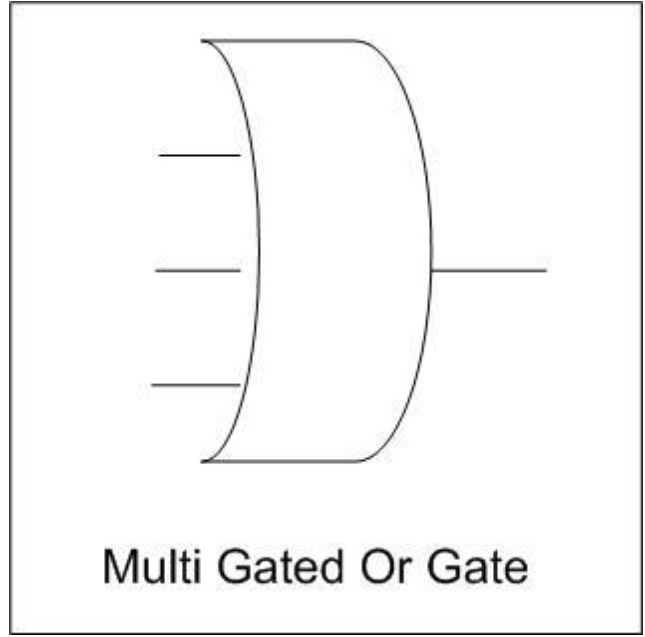

Figure 4.

| P | Q | R | P∨Q∨R |
|---|---|---|-------|
|   |   |   |       |
| T | T | T | T     |
| T | T | F | T     |
| T | F | T | T     |
| F | T | T | T     |
| T | F | F | T     |
| F | T | F | T     |
| F | F | T | T     |
| F | F | F | F     |

Table 2. The corresponding Truth Table for this multi-switched OR gate.

The same being true for a multi gated OR gate. An OR gate with multiple inputs. Not a cascade of OR gates.

Since our technology is currently limited to binary, we can not begin to imagine the possible advances a "multi-branching" technology would have.

[I would posit it is correct to say, that our –or Western—logic is based upon binary as well. That Aristotlean logic, syllogism etc., is based upon binary choices and branching. Going beyond binary logic is to open a door to another dimension in logic.]

I toyed with the idea of a multi-switch that could be implemented with our current technology and interfaced with our current computer architecture.

Originally, I thought an optical solution would be possible. X-ray diffraction through crystals allows for a standard dispersal pattern. From one optical input there could be many outputs.

However, this would have difficulties in implementation. Physical bulk for one.

Another serious processing limitation to crystals is, that the crystals have a fixed pattern. Yes, a particular crystal might disperse an input light signal to 20 points. But, there is no control or selection possible. Perhaps only 10 or 19 switches should be turned on. This might be overcome with filtering or polarization. However, this would be bulky and mechanical.

So, I abandoned x-ray diffraction as a feasible approach.

Then, ancillary to contemplation on approaching a new research project in bio-informatics applied to neuroscience, another idea for an electronic solution came to me. Use multiple voltage inputs with a Zener diode.[1] Let each input, be an equal voltage, who's total will be the threshold voltage needed to fire the Zener diode. This is something which we have the technology to implement; easily interface with current computer architecture and; would provide for a switch that toggles on from more than one input. Also, the technology would provide the capability for multiple programming threads to merge, reconvene—in hardware and software. I.e. Separate programming threads could coordinate processing. (This is discussed in depth, below, with examples.)

The converse use of the technology would provide the capability of spawning multiple processes at once. (This will be explained, in detail, below.)

Also, these ideas led me to an optical solution. Optical technology opens up many new vistas, including greater speed of operation and bandwidth. Such a method would do the job of supra-binary computing quite well. Optical technology would have to interface with digital electronic technology.

I have limited this paper to a discussion of an electronic switch.

A quick analysis of programming constructs reveals that such a procedure would produce RISC programming with tighter code. A significant reduction in logic instructions would be achieved. (This is discussed in depth below, with examples.)

Currently, all computer circuitry is based on 2 voltage levels. Meaning, there is a base voltage and a high voltage that is used to turn on a switch, logic gate. [MLV01] However, we can build switches and logic gates that would need to accumulate more voltage. These mutli-switches would "fire", send a signal, at a higher voltage threshold than a regular switch. See the following figure 5. (This is elaborated upon, below.)

---

[1] To the best of my knowledge, Zener diodes are currently, physically, too big for very large scale miniaturized circuits. Devices for routers and networking can be built. Perhaps, even peripheral devices. However, registers or CPUs, would be too bulky. While not an electronic engineer, I think some kind of miniaturized device or PN junction could be made to perform this function.

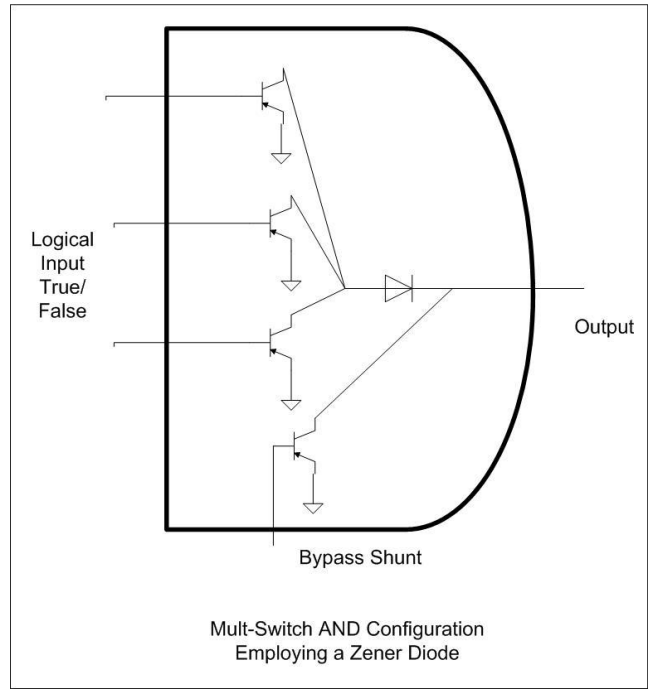

Figure 5.

Another construction, would be a minor tweaking of current technology. A multi gated AND gate could be constructed of several transistors in series. As in the following figure 6.

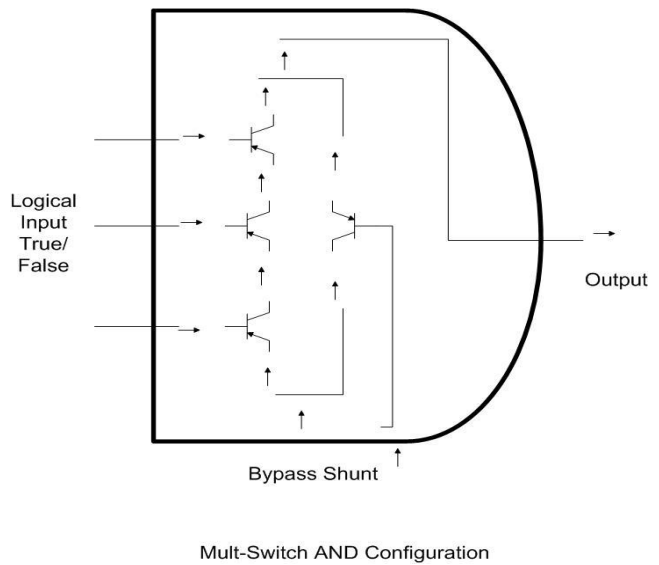

Figure 6.

A multi gated OR gate could be constructed with several transistors in parallel that would turn on based upon any one of the inputs. As in the following figure 7.

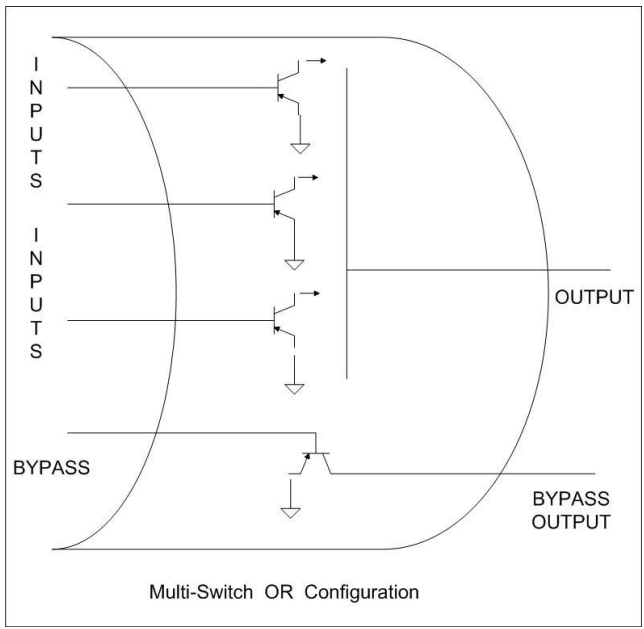

Figure 7.

Then, there are more sophisticated permutations of these possible constructions. We are not limited to 3 or 4 inputs. But, we could build base 10 computers with this design. Even, CPU registers, counters and clocks could be built on this design—a true 12 or 24 hour clock.

Also, the designs shown, while having more than binary input and activate with a combination of voltage levels, the *output* is still binary. These devices could be built to output several voltage levels. This would be a whole other dimension to signal processing and allow the use of bases greater than base 2. This discussion is only going to be mentioned in this article and not going to be covered in depth in this article. [This idea is much more developed and technically easier to implement in a photonic solution.]

Of course, these hardware devices would require special "multi-switch" programming constructs; a specially designed compiler to implement the "mutli-switch" instructions into machine code and; special "multi-switch" hardware.

Of course the number of multi-switches that a program could use, would be limited by the physical constraint of the actual number of hardware switches. But, such limitations are to be expected. After all, there are memory constraints; limitations to stacks, nesting if statements, subroutine calls, etc.

It would also be possible, to save a multi-switch state and load a new state. Just as is done with registers during a subroutine call in machine language. [PET01]

This "paging" or "swapping" would permit an unlimited amount of multi-switches no matter what the hardware constraint. But, the "swapping" would slow down operations. It would also be very time consuming, use a lot of paging, for 2 logical multi-switches committed to the same logical operation; to page and share the same physical multi-switch.

While the applications are as yet unknown, certainly real time processing control systems would benefit from such computing ability. As would: neural networks and database kernels.

Imagine based upon one input, such as from a smoke detector; that would simultaneously—literally at the same time—lockdown airtight a facility, contact the fire department and commence other fire extinguishing measures.

This may appear to happen with current technology, but it does not. It appears so because the time lag is so minute, the time lag does not register to human beings. But, in computing architecture, these alarms are triggered sequentially, one after the other.

Many similar scenarios are possible.

In mission critical applications, the simultaneity of a multi-switched system would be a plus.

Although possible hardware implementations are discussed, this paper is purely theoretical. The black box approach has been taken. Predicated upon the assumption that the technology is possible, what-if scenarios about the technology are discussed. A prototype was not yet been built.

In the following order, this paper will discuss:

- An explanation as to why multi-switching is a RISC method of computing.

- A new hypothetical programming construct called "multi-switching".

- A discussion of necessary additions to a compiler for a programming language with multi-switching.

- Some sample implementations of the multi-switch construct.

- A theoretical discussion of possible hardware implementations.

## 2. MULTI-SWITCH PROGRAMMING IS RISC PROGRAMMING

Reduced instruction sets allow for maximum speed of execution, with tight code. [CLE01] (Although RISC programming is usually applied to *microprogramming*, the programming of microprocessors, the advantages of RISC programming can be applied to any programming task. Good engineering, including software engineering, includes simplicity – minimizing complexity. Unnecessary complexity should be avoided. [MCC01] [IEE01] Also, the greatest advantages of RISC, come with moving data. [CLE01]

As there would be both a high level and low level multi-switch function, the RISC character of multi-switching would apply both to high level programming and microprogramming.

This simplest and basic example is from an if…then construct.

Begin with a simple example of instruction reduction:

if (a == x) then *action #1*

if (a == x ) then *action #2*

if (a == x) then *action #3*

This is 3 statements, 3 comparisons and 3 actions.

Clearly, this is tighter, less code, more efficient when written and programmed as:

if (a == x) then  {

    *action #1;*

    *action #2;*

    *action #3;*

    }

This is 1 statement, 1 comparison and 3 actions. The time taken in loading each comparison separately is avoided. Along with reducing the number of comparisons, the *time* in making those comparisons is reduced. There is after all, only one comparison made instead of three.

However, to execute all three statements, requires loading each statement individually and sequentially. Then, executing each statement. What if each statement were loaded and all *three* were executed, *at once*, if the logic test proved true? This is the advantage of parallel processing. What if *three* branches in code were executed at once? Then, 3 statements are executed in the time of one. If 10, then an order of magnitude in processing is achieved.

Now, consider the following code:

if (a == x) then  {

    if (b == y ) then  {

        if (c == z) then *action*

    }

}

This is 3 statements, 3 comparisons and 3 actions. Each true comparison requires a statement to fall through to, to the subsequent code in the nesting structure—which is a new if…then with its own, new comparison.

The code is tighter and has less instructions if rewritten as:

if ( (a == x) and (b < y ) and (c > z) ) then  *action*

Now, there is 1 statement, 3 comparisons and 1 action.

Each comparison requires the loading of the comparison operands; a comparison; counting a system variable or setting a system flag and; if the comparison is true, continuation down the line to the next comparison. If all *three* comparisons are true, *then* a branch is executed to the action code.

What if all 3 comparisons are executed at once? Then, 3 comparisons are executed in the time of one. What if no system flag or counter is necessary to register the validity of multiple comparisons?

The more comparisons, the greater the savings in time, as more comparisons are compacted together into one comparison. This is RISC programming.

How many complex comparisons are made in the average program? How complex are these comparisons? This requires further study. How much savings could be made by the implementation of such a technology? That would depend upon the outcome of a study of how many complex comparisons are made in the average program and; how complex are the comparisons.

## 2.1 MULTI-SWITCH PROGRAMMING IS PARALLEL PROCESSING

Take the following if…then statement:

if (a == x) then  {

    *action #1;*

    *action #2;*

    *action #3;*

    }

Current technology implements this with one comparison and *three* separate executions of code that are executed sequentially. However, a multi-switch could branch to three separate locations—at once. This is *natural* parallel processing.

The following schematics demonstrates a multi-switch with multiple outputs. Either, the outputs could be polled for activating a process or programming thread. Or, the multi-switches would have their outputs tied to address registers that initiate threads. The contents of these registers would indicate the next instruction of program code to execute. Or, the outputs could be hardwired to activate a piece of hardware to initiate a process.

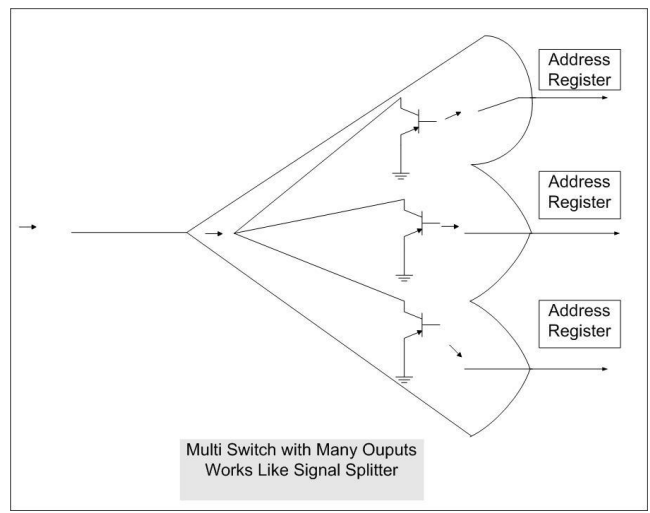

Multi Switch with Many Ouputs Works Like Signal Splitter

Figure 8.

In the case of inputs, from programming threads; the inputs would accrue, in a multi-switch, until all threads had completed and the next stage of programming could be initiated.

This fulfills a requirement of parallel processing to wait for threads to complete. [DEI01]

See Figures 5 & 6 above.

The bypass shunt would be used for time-outs for threads that were taking an inordinate amount of time to complete. This might indicate, for example, that a thread had gone into an infinite loop. In which case, a bypass would be necessary to prevent an interminable wait.

Of course parallel processing would require more than one CPU. But, multi-switching makes parallel processing innate. Coupled with an address, in a separate program counter register, a multi-switch output would turn on a thread or initiate a process in a new or alternate CPU, with code located in a specified address.

## 3. THE MULTI-SWITCH PROGRAMMING CONSTRUCT

The multi-switch programming construct would take several forms. There would be a machine level multi-switch construct and a high level multi-switch construct. Either way, the multi-switch construct would contain a table of targets, branches, addresses for branching. These branches might be physical addresses or labels for a "goto".

Also, there should be a reset function, that zeroes out the branching addresses. This would be good for initialization and other operations.

The number of possible branches that the switch can go to, I will call "the size" of the multi-switch. E.g. If a multi-switch can make 3 branches, the "size" of the multi-switch is 3.

Here are a few examples:

I will call the multi-switch operator "mswitch".

I will call the multi-switch reset operator "msreset".

The syntax:

mswitch *name*(*target, target, target[parameter, parameter...] ...*)

*name*—is the name a particular mswitch

*targets*—can be either a goto label, a subroutine label, or a process to initiate.

Subroutines and processes may be passed parameters.

Parameters can be returned with their values altered.

There are return codes for subroutines or processes called.

msreset(*name*)

*name*—identifier of a previously declared mswitch, or a fixed name for a specific multi-switch, piece of hardware

Examples:

declare mswitch ms1;

.if(a==x) then ms1(process-a, process-b, process-c)

Example 1.

Initiating the 3 subroutines at once.

declare mswitch ms1;

if(a==x) then ms1(print(form-a), print(form-b), print(from-c))

Example 2.

Initiating the same subroutine 3 times with different parameters.

declare mswitch ms1;

if(a==x) then
        ms1(print(form-a[dept-a prtr, name, id#]),
        print(form-b[dept-b prtr, name, id#]),
        print(from-c[dept-c prtr, name, id#]))

Example 3.

Initiating the same subroutine 3 times with different target output devices.

declare mswitch ms1;

ms1(db-search-a[ id#]), (db-search-b[ id#])

Example 4.

An example of parallel processing.

declare mswitch ms1;

if((ms1(db-search-a[ id#]), (db-search-b[ id#]))==TRUE)
        then machine-found(id#)

else
        print("One or more databases inaccessible. \
        Query not completed.");

Example 5.

An example of parallel processing and using return codes.

The return code is a single return code for the multi-switch operation, applying to all processes called.

All processes called have completed successfully.

Alternatively:

declare mswitch ms1;

ms1(db-search-a[ id#]), (db-search-b[ id#]);

if((ms1.db-search-a)==TRUE) OR  (ms1.db-search-b)==TRUE)

    then match-found(id#)

Example 6.

An example of parallel processing and using return codes.

The return code of each process called is evaluated for successful completion.

declare mswitch ms1;

ms1(db-search-a[ id#, name]), (db-search-b[ id#, address]);

print(ms1.db-search-a.name, ms1.db-search-b.address);

Example 7.

An example of parallel processing and returning parameters.

Alternatively:

declare mswitch ms1;

ms1(db-search-a[ id#, name]), (db-search-b[ id#, address]);

stage-name= ms1.db-search-a.name;

agent-address= ms1.db-search-b.address;

print(stage-name, agent-address);

Example 8.

An example of parallel processing and returning parameters.

declare mswitch ms1;

msreset(ms1)

Example 9.

An example of initializing a multi-switch. Good programming practice.

declare mswitch ms1;

ms1(db-search-a[ id#, name]), (db-search-b[ id#, name]);

msreset(ms1);

ms1(db-search-c[ id#, name]), (db-search-d[ id#, name]);

Example 10.

An example of clearing out the addresses in a multi-switch and reusing the declared multi-switch variable name. Notice the 2$^{nd}$ search looks into different databases than the 1$^{st}$ search.

This example also subtly demonstrates the size limitations of a multi-switch.

# 4. COMPILER CONSIDERATIONS FOR MULTI-SWITCH PROGRAMMING

Some idea about the hardware implementation is necessary to understand the requirements of compiler implementations of a multi-switch. The switch would have to have built into itself or; have access to, a table of addresses for the targets to activate if the switch is turned on. Also, the switch will have to have multiple addressable inputs, the combination of which, will turn on the switch. Knowing these considerations necessitates certain compiler outputs.

The high level language programming construct described, would have to be implemented in machine language. This implementation would have to access the new hardware that would be the multi-switch. Also, the low level language implementation would have to include all the features of the construct described above.

This means the opcode will have to load all the target branching addresses into the multi-switch logic gate. Loading the addresses into a multi-switch gate may require more than one opcode. Because, opcodes are limited to the number of addresses they can contain.

In addition, as the computer has the capability of multi-switching, this capability should be taken advantage of. Many functions should be implemented using the multi-switching technology and function.

An example:

The compiler will decide whether or not to use a multi-switch to implement an if…then statement. This decision would be based upon the number of comparisons needed. Instead of implementing the if…then statement by sequentially loading comparisons and their corresponding branch statements; the compiler would choose a multi-switch construct of many comparisons at once and branch, if all processes complete successfully.

Also, the compiler would have to decide, if the logic requires a capacity of branching that would exceed the mutli-switch. Then, an alternative would be used. Either reverting to old form software imitation of multi-switching or; paging with a multi-switch or; using several multi-switches together, at once, to complete a complicated "if" statement.

## 4.1 Opcodes for Multi-Switching

There will have to be several opcodes for implementing multi-switching:

1. An opcode to initialize the switch, reset all the pointers to target branches. Let us call this opcode "Op-Reset".

2. An opcode to set up the switch *to be activated* if processes, threads or subroutines complete successfully or; if logical comparisons, signals or alarms are true. In other words, an opcode to send an input signal to a specific multi-switch. Let us arbitrarily call this opcode "Op-Turn-On".

3. An opcode to indicate that an input line to the multi-switch is turned off--not that the input process has not

been completed yet rather, completed. For example, a "FALSE" comparison. Let us arbitrarily call this opcode "Op-Turn-Off".

4. An opcode to set up the multi-switch (as shown in figure 8) *to activate* processes or subroutines or program threads or branch to program locations. Let us arbitrarily call this opcode "Op-Activate".

### *4.1.1 Op-Reset*

Op-Reset will zero out all input and target addresses of the multi-switch.

In the high level language implementation of a multi-switch, when a multi-switch label is declared, the compiler should (good programming and software engineering design) clear out all input and target addresses in the multi-switch. This would mean a call to Op-Reset. Only after clearing out the input addresses, then the input addresses should be assigned.

### *4.1.2 Op-Turn-On*

Op-Turn-On will have to have the addresses of the inputs to the multi-switch logic gate. Note there are several inputs to a multi-switch.

Different processes or comparisons will have to access different addressable input lines to the multi-switch gate. Yet, there is only one physical and logical switch.

The compiler would have to keep track of the physical address of the multi-switch and the physical addresses of the input lines to the multi-switch.

Also, the compiler will need a table to keep track of the assignment of input lines. Before assigning a new input, the compiler will have to know which lines are available and; which lines have already been assigned. The compiler will have to have address resolution that will assign the logical input lines to the physical input lines of a switch.

### *4.1.3 Op-Turn-Off*

During operation, when a multi-switch is used, it will be activated by several input lines. This will be done by sending a "high" or "on" signal. Otherwise, the signal will be "off". The signal is turned "on" when a process completes successfully with a desired result or; a comparison is made and evaluates true, etc. In the scenario when the process has completed *unsuccessfully* or the comparison is *false*, no signal will be sent to the multi-switch to activate it. However, the multi-switch needs to know this too. Otherwise, the multi-switch will wait, in limbo, forever, for all input lines to turn on.

This may be resolved with a timing solution or; a "completion table" of completed tasks (comparisons, processes, subroutines, etc.)

It may also be resolved by sending input signals, to "bypass", shunt the multi-switch. This would be the preferred method. It is simple and would indicate a failed or false operation.

Using a time limit for processes to complete may give erroneous information. Some processes may just take longer to complete.

There will be the issue of how to handle infinite loops. What if a multi-switch is waiting for a signal from a process in an infinite loop? My personal recommendation would be to handle such problems with standard debugging techniques—not to implement a timeout. However, a timeout is used in some operating systems to avoid infinite loops. [DEI01]

A bypass mechanism would be a better and simpler solution, in my personal opinion and; involves a lot less operations than maintaining a completion table while continually parsing the table for results. However, a table would give a multi-switch the ability to vary in "size".

Since the multi-switch is designed to be turned on when all inputs are high, one single bypass signal / line should be sufficient.

The opcode op-turn-off would send a signal to the shunt, turning the shunt on, which would turn the multi-switch "off". The multi-switch will send out a low voltage or "off" signal.

The opcode op-turn-off would have be added to every process or comparison that would turn the multi-switch on. The op-turn-off would be activated by an if statement. If the process fails or the comparison is untrue, then op-turn-off is activated.

The opcode op-turn-off must contain the address of the bypass line of the multi-switch.

### *4.1.4 Op-Activate*

Op-Activate will have to have the addresses of the output lines from the multi-switch logic gate. Note there are several outputs to a multi-swich.

Different processes or comparisons will be activated, branched to, from different addressable output lines from the multi-switch gate. Yet, there is only one physical and logical switch. But, there are several output lines to that switch.

The compiler would have to keep track of the physical address of the multi-switch and the physical addresses of the output lines from the multi-switch.

Also, the compiler will need a table to keep track of the assignment of output lines. Before assigning a new output, the compiler will have to know which lines are available and; which lines have already been assigned. The compiler will have to have address resolution that will assign the logical output lines to the physical output lines of a switch.

## 5. SAMPLE SOFTWARE IMPLEMENTATIONS OF THE MULTI-SWITCH PROGRAMMING CONSTRUCT

### 5.1 If…Then

A complex if…then statement would be implemented in a machine level multi-switch construct that would function similarly to the high level construct in (programming) Example 5. The comparison part of the if…then statement would be broken up into parts. Each comparison would be treated as a process. The comparison processes would execute in parallel. As each process completes, it would send a logical yes to an input line of a multi-switch gate. The multi-switch gate will go on only if all processes complete successfully.

### 5.1.1  Parallel Processing

As demonstrated with the example of an If…Then construct, the multi-switch is a hardware method of implementing parallel processing. This has an advantage over the current software implementation. [DEI01] Current implementation of parallel processing separates program code into threads. When a program comes to a point where it branches to separate threads, it runs each separate thread. Later, the program continues after all the threads have terminated. As separate threads are executed and terminated, it is necessary to know that all the threads have completed, before the program can continue. Currently, this is done with semaphores. Again, **semaphores are a binary device.**

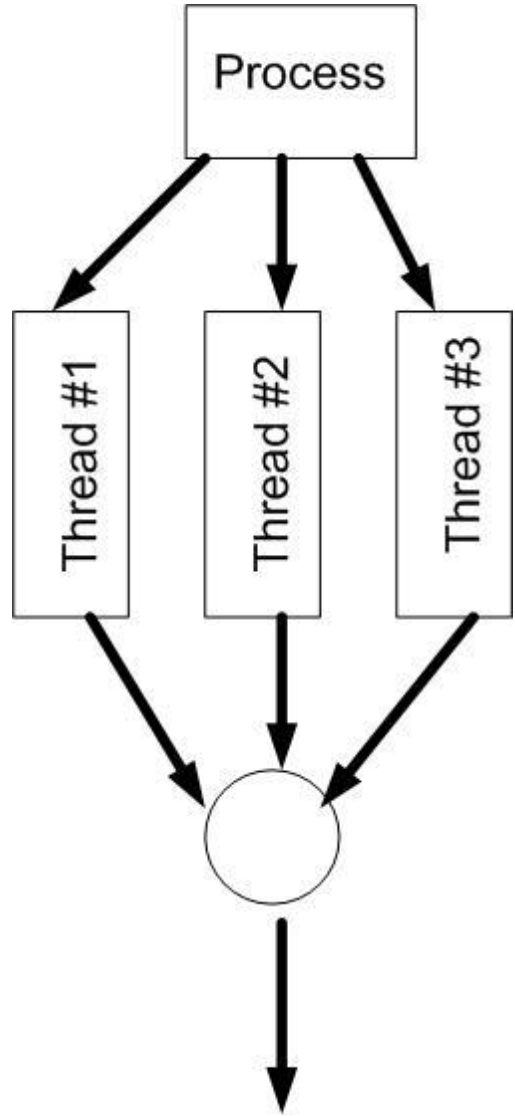

Figure 9.  Basic Thread Processing

Each thread needs a separate semaphore. Each semaphore has to be monitored. All the semaphores, together, for all the threads, must be polled, continually; until, all the threads complete. A time consuming process.

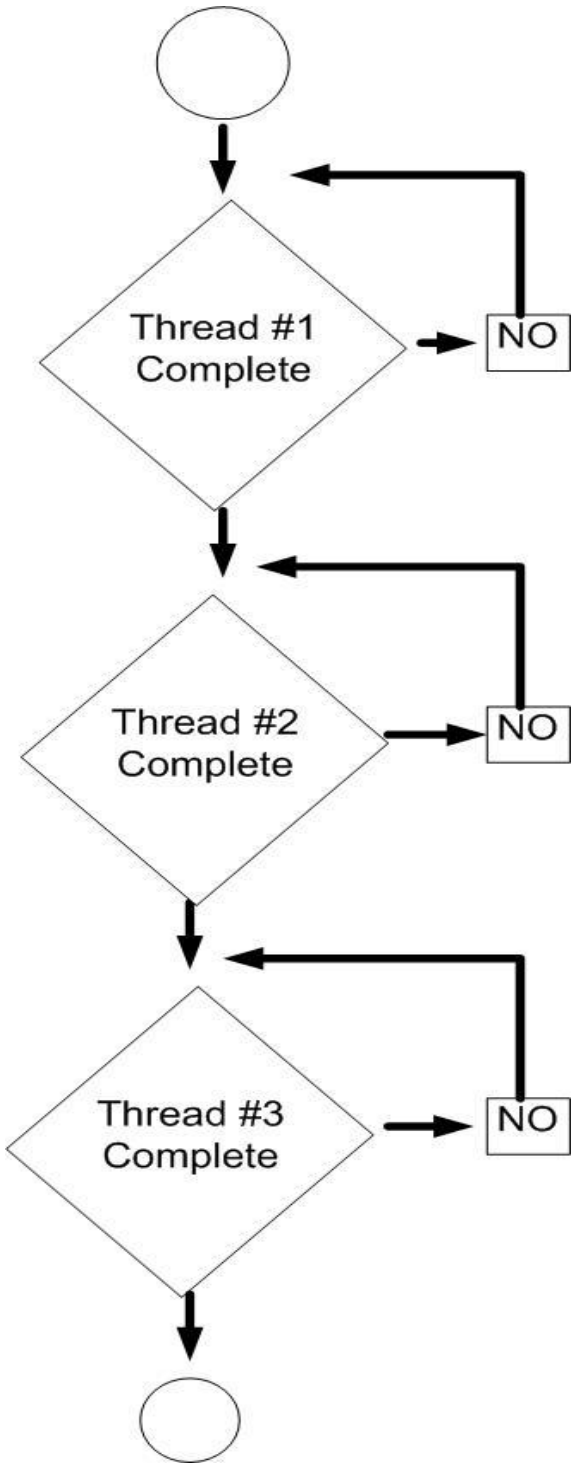

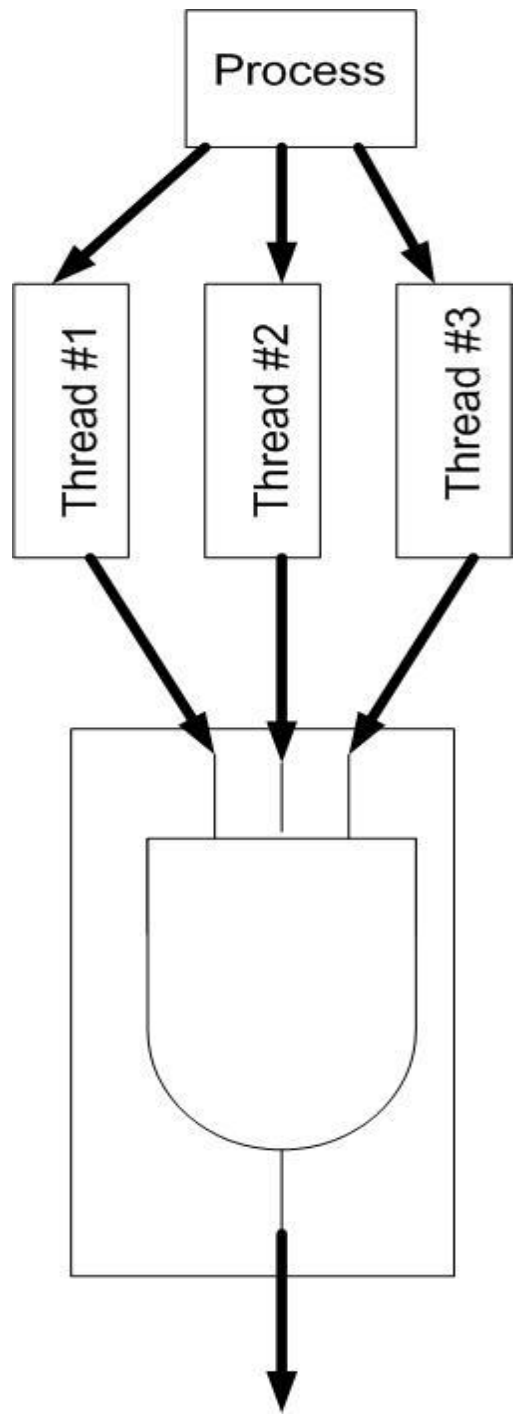

Figure 10. Thread Completion Process Flowchart

A multi-switch would provide the function of determining if all the threads have completed, in less time, without constant polling. As each thread completes, it would send a signal to a multi-switch. The multi-switch will fire when all threads have completed. The multi-switch will then signal the program to continue execution from the recombination program point.

Figure 11. Using a Multi-Switch to Complete Parallel Processes

This is a much simpler and much more direct approach to parallel programming. Direct approaches are better, especially when eliminating unnecessary complexity.

The number of threads that can be handled would depend upon the size of the multi-switch.

One implementation of parallel processing would use several multi-switches, each of a different size. The compiler would select the right size multi-switch for the particular number of threads in an instance of multi-programming.

However, one multi-switch with many input lines could handle any number of threads less than its "size"—number of input lines. This could be easily programmed. Ex. A multi-switch has 10 input lines. There are 3 threads. The multi-switch can handle that.

For parallel processing, I foresee the multi-switch as an excellent advancement in hardware. It will provide efficiency, speed, and reduce programming code. All in all, it will simplify and streamline parallel processing as well as making parallel programming a much more practical way of programming.

## 5.2 Implementing a neural net with a multi-switch.

A neural net is currently implemented as a hierarchy of nodes. Each nodes points to successive nodes in subsequent levels, beneath itself. This is similar to an inverted pyramid and top down approach. However, a pyramid starts at one point. A true neural net, starts with many nodes at the top. Current designs of neural networks do not allow for lateral and upwards reference. While these may be logically more streamlined and, perhaps easier to program, it may not be real world.[2] [BAR01] [&&&QUI01 locust paper]

In fact, the lack of lateral and upwards links in current neural network implementations is a shortcoming of our neural networks. For neural networks to be optimal, they should include lateral and upwards links. Without lateral and upwards links, the neural network is rigid and unable to reconnect. It is partially static and limited. It is not fully dynamic and free to choose links as necessary.

Ex. In a real neural network, a fact, that might have been considered ancillary and at the bottom of the pyramid can be relocated to the top of the pyramid.

Real neural networks can reconstruct themselves. This is something current neural networks do not really do.

However, a neural network built with multi-switches can reconstruct itself.

In a neural network, with multi-switches, a multi-switch would itself be a node.

It would address the input lines of other multi-switches.

This would allow for a node to point to other nodes. However, the top down hierarchy would not have to be maintained.

---

[2] Natural networks have conventional links. But, they also have "shortcuts". [BAR01] Natural neural networks have latitudinal *as well as* longitudinal connections. When this article was originally conceived, latitudinal connections were not used. Technology has evolved, [HAG01] [QUI01] The "multi-switches" herein conceived would allow for physical and logical neural networks with latitudinal, longitudinal, upwards and downwards connections as well as shortcuts.

Also, an application such as a neural network, could use many multi-switches. In fact, specific machines, rich in multi-switches might be designed especially for neural networks.

A neural network would be an application that would make good use of saving and reloading logical configurations (symbolic addressing) of multi-switches. As the number of multi-switches might be less than the size of the neural network, the logical design of multi-switches, "which nodes link to which nodes", could be preserved; by saving the software assignment—symbolic addressing of a multi-switch; swapping out the multi-switch's configuration (symbolic addressing) with a new configuration and; restoring an old multi-switch configuration as needed.

A configuration containing the output addresses of a multi-switch. In this case, that would be symbolic multi-switch names and not physical addresses of multi-switches.

## 5.3 Implementing an indexed database with a multi-switch.

Many databases maintain indexes dynamically. Databases are frequently indexed by different keys. Using a multi-switch design, which would implement parallel processing, a record could easily be added to a database and a procedure to each index would update all indexes instantaneously.

Likewise, the reverse could be done when deleting a record. It could be removed from all indexes simultaneously.

## 5.4 Implementing a database search with a multi-switch.

A multi-switch design would be of advantage when searching for complex keys [a key made up of a combination of several data fields or parts of data fields] or; the equivalent of a complex key, with the data fields, but an actual index has not been made for such a complex key.

Using a multi-switch and parallel processing, a search could be initiated to look for those records with matching fields. When all the fields match, i.e. the components of the key, the desired record would be found.

This would be different than currently looking for the primary field. Finding a match then, inspecting the secondary record. And so on, until the complete key is found.

This multi-switching technique would check all relevant fields in a database simultaneously.

Multi-switched computers would reduce search time by combining search commands and; by turning multiple searches into single searches for multiple items.

## 6. POSSIBLE HARDWARE IMPLEMENTATIONS OF A "MULTI-SWITCH" COMPUTER LOGIC GATE

There are two possible implementations I have thought of. One is electronic. The other is optical. This paper will discuss an electronic implementation only.

## 6.1 An Electronic Switch

The electronic switch can be implemented in either a hardwired of soft set manner. I will discuss both.

### 6.1.1 Hardwired

A hardwired solution can be implemented. I have thought of several possible implementations: Multiple inputs to activate a higher threshold logic gate (like a Zener diode); transistor gain with multiple inputs to activate the switch and; a shunt to bypass the switch if the comparisons or processes fail.

See figures 5, 6 & 7.

#### 6.1.1.1 Multiple Port Logic Gate

A logic gate can be built with 3 or more input lines. The logic gate would work at the same voltage levels that are currently used. The trick would be that the switch will be built with a material like a Zener diode. Zener diodes have a high resistance. But, after a threshold is reached, the resistance barrier is broken and the current flows. The threshold voltage will be set high enough that it will require the combined voltage of all the input lines.

Alternatively to having more than 2 voltage levels, the Zener diode could be selected, or put in series with a resistor, so that the output voltage would be at the same level as an ordinary high voltage signal.

See figure 5.

#### 6.1.1.2 Transistor Gain Increments To Higher Voltage Level Input Threshold

A logic gate can be built with 3 or more input lines. The logic gate would work at the same voltage levels that are currently used. The flow through the switch will be controlled by transistors. The transistors would be in series. The gain will be controlled by the combination of all the input lines. The transistor will be selected so that a high (on) output signal would only occur with the combined voltage of all input signals. (Similar to current AND gate operation. Just more inputs.)

See figure 6.

#### 6.1.1.3 Shunt for bypass, if 1 many tests fails.

The multi-switch will have a bypass input activated by comparisons or processes, as they complete with a "FALSE" or failure return code. A special opcode, op-turn-off, will be executed at the end of the comparisons or processes, to send a signal to the bypass line. The only high signal coming from the multi-switch will be from the shunt/bypass, indicating a logical FALSE.

See figures 5, 6 and 7. Note the bypass input and output lines.

### 6.2 Soft Set Channel Table

A multi-switch with multiple outputs (See figure 8.) will need a table of some kind, to indicate where the outputs should go. Meaning, the addresses of the output locations associated with the output lines.

This could be implemented in either a hardwired or soft coded fashion. Either the target addresses could be stored in special registers—that might be built into the multi-switch. Or, the target addresses could be stored in a table in memory, that would be accessed as needed [when an output line goes high].

Using a table instead of hardwiring outputs would require more memory accesses; which might slow down operation. However, using a table in memory would make it easier to swap logical multi-switches amongst physical multi-switches.

### 6.3 When activated, the multi-switch should know if there are no more targets to go to.

In figure 8, the multi-switch has 3 outputs. The multi-switch appears "static" in that it must go to all 3 outputs—no more, no less.

However, as stated in the beginning of this paper, the ability to scale—chose the number of paths a switch should have, was a design consideration.

While I can conceive of a switch with a potential range of up to 20 outputs—as the example from a crystal's X-ray diffraction, I doubt that is practical. However, having a need to be able to scale from 3 to 5 switches does have an intuitive reality.

How would the scaling be done?

To use only some of the lines, it would be necessary to have an indicator, that there is no address to go to. For example, a known bad address. All zeroes is a common way of doing this. If the address in the address register points to all zeroes, the switch & compiler, know that that line is not being used.

If the multi-switched is hardwired, the address register would have to an invalid address. Instead of causing an exception, the invalid address and output signal, should simply be ignored. No signal should be sent to that address.

If the multi-switch is soft coded with a table, the compiler should simply not generate opcodes for output lines that have invalid addresses.

### 7. FUTURE DEVELOPMENT

### 7.1 Study the number of multi-switches current programming would use.

Actual programs, a large sample of code, will need to be analyzed to determine the number of switches average multi-switches will need to satisfy current programming needs.

Also, an appropriate number of multi-switches per computer will need assessment.

While a multi-switch of size 3 will take us beyond binary, larger switches will probably be of value. However, multi-switches of size 20 will probably be a waste, unless is special cases, perhaps telecommunications. For math, I am confident there is a desire to produce real base 10 computers. Hence, 10 line multi-switches would be necessary.

The matter needs study. We have to look for the answer to questions like:

- How many concurrent branches are found in existing programs?
- How many branches do concurrent branches have, in existing programs?
- What is the largest number of concurrent branches in existing programs?

- What is the average number of concurrent branches in existing programs?

The answers to these questions will guide us in building multi-switched computers. These answers will tell us how many multi-switches to build and; what switching size these switches should have.

I predict, that with time and popularity, the number of multi-switches built into a machine, will increase.

## 7.2 Prototypes Under Consideration

The construction of a demo electronic multi-switch.

The construction of a demo optical multi-switch.

The methodology employed in the optical switch would increase the bandwidth of current technology by a minimum of ninefold (9x). (With current technological capabilities.)

This could be applied in a downwards compatible method. Meaning, current cabling infrastructure would not need alterations.

## 8. SUMMARY

Author's Bio: Givon Zirkind received his Bachelor's in Computer Science from Touro College and; his Master's in Computer Science from Fairleigh Dickinson University, both schools are located in the USA. His career has involved computer operations; design and management of business applications with extensive database programming and management; Internet, computer communications, data transfers and telecommunications; data conversion projects; reverse engineering of data and legacy software; being a published author and editor of a technical journal as well as a technical writer; teaching and; automated office support. In addition to his AFIS data compression work; he has done independent genetic database development and research. Also, he has developed non-decryptable encryption. (See article in the Archiv database at Cornell University.) He may be reached at his email: GIVONNE@GMX.COM.

## 9. ACKNOWLEDGMENTS

To my grandfather, R.I.P., for all his support in all my endeavors.

To Ramona Brandt for her support to this project.

Dr. Larry T. Ray, Ph.D. R.I.P., Mathematics/Computer Science, Stevens Institute of Technology (NJ), formerly professor of computer science, Fairleigh Dickinson University, for his support in my projects.

To Mrs. Ray, for her continued support.

To Dr. Gertrude Levine, Fairleigh Dickinson University, for her interest in my work and her referral that led me to this project.

To Dr. Alexander Casti, Fairleigh Dickinson University, who introduced me to Dr. Ehud Kaplan and Dr. Kaplan's research. Without that introduction, I would have never been included in this project.

To Dr. Ehud Kaplan, Mt. Sinai School of Medicine, for offering the opportunity to be included in his research, access to his data, the opportunity to donate graphics & video card programming and; the opportunity to mine his data any way I saw fit.

To Carol Crystal, friend and colleague. For her support in this and other projects.

To Allen Gerner, friend and colleague. For proofreading and support.

To Ralph Yozzo, friend and colleague. For his support.

## 10. Disclaimer

The ideas and views expressed in this paper are my own and do not reflect the views of the Mt. Sinai School of Medicine.

_________________________________________